\documentclass[sn-mathphys-num]{sn-jnl}

\usepackage{graphicx}%
\usepackage{multirow}%
\usepackage{amsmath,amssymb,amsfonts}%
\usepackage{amsthm}%
\usepackage{mathrsfs}%
\usepackage[title]{appendix}%
\usepackage{xcolor}%
\usepackage{textcomp}%
\usepackage{manyfoot}%
\usepackage{booktabs}%
\usepackage{algorithm}%
\usepackage{algorithmicx}%
\usepackage{algpseudocode}%
\usepackage{listings}%

\begin{document}

\title{Three-dimensional repulsive Hubbard model}%

\author{\fnm{Alexei} \sur{Sherman}}\email{alexeisherman@gmail.com}

\affil{\orgdiv{Institute of Physics}, \orgname{University of Tartu}, \orgaddress{\street{1 W. Ostwaldi St.}, \city{Tartu}, \postcode{50411}, \country{Estonia}}}

\abstract{The three-dimensional repulsive Hubbard model is investigated using the strong-coupling diagram technique. For half-filling, the boundary between paramagnetic and antiferromagnetic states is determined for the range of the Hubbard repulsion $4t\leq U\leq12t$, where $t$ is the hopping integral between neighboring sites. Along this boundary, the density of states is calculated, and it demonstrates the Mott transition at $U\approx9t$. For $U\geq6t$ and half-filling, the density of states has the shape inherent in the strong electron repulsion, while for $U=4t$, its shape points to weak coupling. The dependence of the N\'eel temperature $T_{\rm N}$ on the electron concentration $\bar{n}$ is investigated for the cases $U=4t$ and $12t$. In the former case, $T_{\rm N}$ decreases monotonously with $\bar{n}$, while in the latter case, there is a plateau in the dependence near $\bar{n}=0.87$. The plateau is connected to a reconstruction of the density of states caused by an effective weakening of electron coupling due to electron depopulation. For $U=12t$ and half-filling, the magnetic critical exponent $\gamma\approx1.4$, which is close to the value in the Heisenberg model. Some features resembling the first-order phase transition, revealing themselves in a finite crystal, are discussed.}

\keywords{three-dimensional Hubbard model, N\'eel temperature, density of electron states, Mott transition, magnetic critical exponent}

\maketitle

\section{Introduction}
The interest in the three-dimensional (3D) repulsive Hubbard model is connected with its rich phase diagram, which contains regions of the Mott insulator, bad metal, Slater gap, and para- and antiferromagnetism. The model was investigated with Monte Carlo simulations \cite{Hirsch,Scalettar,Staudt}, the dynamic mean-field approximation (DMFA) \cite{Jarrell,Ulmke}, its cluster \cite{Kent} and diagrammatic \cite{Rohringer,Hirschmeier} extensions, and the determinantal diagrammatic Monte Carlo \cite{Kozik}. As a results of this work, the dependencies of the N\'eel temperature $T_{\rm N}$ on the Hubbard repulsion and electron concentration $\bar{n}$ were obtained. Data provided by the above methods give a rather wide scatter in $T_{\rm N}$ for values of the Hubbard repulsion $U\gtrsim8t$, where $t$ is the hopping integral between neighboring sites. Therefore, it is of interest to investigate this region with the method especially developed for the strong coupling -- the strong coupling diagram technique (SCDT) \cite{Vladimir,Metzner,Pairault,Sherman18}. The approach has a number of advantages in comparison with the above diagram methods. In its framework, one is able to describe the Mott transition, while the mentioned methods use the weak coupling diagram technique, which in itself cannot do this \cite{Moukouri}. To represent this transition and the strong-coupling part of the phase diagram, these methods apply data from DMFA, which belong to the infinite-dimensional case. Another merit of SCDT is the possibility to solve exactly the Bethe-Salpeter equations (BSE) for four-leg vertices defining the spin susceptibility. This fact allows one to describe its behavior in the fluctuation region of the phase transition with high accuracy.

Carrying out calculations in 8$^3$ and 10$^3$ lattices, for half-filling, we found the phase boundary $T_{\rm N}(U)$ for the range of the Hubbard repulsion $4t\leq U\leq 12t$. For large $U$, the boundary tends to the dependence $T_{\rm N}=3.83t^2/U$ known from the high-temperature series \cite{Rushbrooke} and Monte Carlo results \cite{Sandvik} in the spin-$\frac{1}{2}$ Heisenberg model. For $U\geq6t$, our obtained values of $T_{\rm N}$ are close to those derived in Monte Carlo simulations \cite{Staudt}. The calculated density of states (DOS) at $U=4t$ is inherent in weak electron correlations, while DOSs for larger repulsions are characteristic of strong coupling. On the phase boundary, the Mott gap opens in the DOS for $U\gtrsim9t$. With doping, the N\'eel temperature decreases differently in the weak, $U=4t$, and strong, $U=12t$, cases. In the former case, it is monotonous, while in the latter, there is a plateau in the dependence $T_{\rm N}(\bar{n})$ near $\bar{n}=0.87$. The evolution of DOSs with doping allows one to understand the origin of this plateau. Depletion of the electron concentration makes the double site occupancy increasingly scarce, which is equivalent to the interaction weakening. In the DOS shape for $U=12t$, this fact is reflected in the change of its character from strong to weak coupling type near $\bar{n}=0.87$. For $U=12t$ and half-filling, the critical exponent derived from the calculated static magnetic susceptibility at the antiferromagnetic momentum is approximately equal to 1.4, which is close to the value obtained in the Heisenberg model \cite{Holm}. Analysing the fluctuation region in a finite-size crystal, we found some features of the first-order phase transition, which weaken as the crystal size grows.

The article is organized as follows: The model Hamiltonian and the main formulas are given in the next section. The results of calculations and their discussion are brought up in Section 3. The last section is devoted to concluding remarks.

\section{Model and SCDT}
The Hamiltonian of the 3D fermionic Hubbard model reads
\begin{equation}\label{Hubbard}
H=\sum_{{\bf ll'}\sigma}t_{\bf ll'}a^\dagger_{{\bf l'}\sigma}a_{{\bf l}\sigma}+U\sum_{\bf l}n_{\bf l\uparrow}n_{\bf l\downarrow},
\end{equation}
where vectors ${\bf l}$ and ${\bf l'}$ label sites of the simple cubic lattice, $\sigma=\uparrow,\downarrow$ or $\pm 1$ is the spin projection, $a^\dagger_{\bf l\sigma}$ and $a_{\bf l\sigma}$ are electron creation and destruction operators, $n_{{\bf l}\sigma}=a^\dagger_{{\bf l}\sigma}a_{{\bf l}\sigma}$ is the site occupation operator, $t_{\bf ll'}$ is the hopping integral. In this work, only the nearest neighbor hopping integral $t$ is taken to be nonzero.

As mentioned above, in calculating Green's functions, we use the SCDT \cite{Vladimir,Metzner,Pairault,Sherman18}. The method is intended for the case $U\gg t$ and uses the series expansion of Green's functions in powers of the kinetic energy. Terms of the expansion are products of the hopping integrals and on-site cumulants \cite{Kubo}. As in the weak coupling diagram technique \cite{Abrikosov}, the linked-cluster theorem is valid, and partial summations are allowed in SCDT. Earlier this approach was applied to the two-dimensional (2D) Hubbard model on a square \cite{Sherman18,Sherman21} and triangular \cite{Sherman24} lattices, the Emery model \cite{Sherman20a}, the extended Hubbard model \cite{Sherman23a,Sherman23b}, and the Hubbard-Kanamori model \cite{Sherman20b}. Results obtained in these works are in good agreement with the available Monte Carlo data.

The terms of the SCDT expansion can be visualized by depicting $t_{\bf ll'}$ as directed lines and cumulants as circles, with the number of outgoing and incoming lines equal to the number of creation and destruction operators in them. Since we are interested in the magnetic properties of the considered model, we sum up an infinite series of ladder diagram describing interactions of electrons with charge and spin excitations. For completeness, main formulas used in the calculations are given in this section. Their derivation can be found in Ref.~\cite{Sherman18}. We name a two-leg diagram irreducible if it cannot be divided into two disconnected parts by cutting a hopping line $t_{\bf l'l}$. Denoting the sum of all such diagrams by $K$, the Fourier transform of the electron Green's function $G({\bf l'}\tau',{\bf l}\tau)=\langle{\cal T}\bar{a}_{{\bf l'}\sigma}(\tau')a_{{\bf l}\sigma}(\tau)\rangle$ is written as
\begin{equation}\label{Larkin}
G({\bf k},j)=\left\{\left[K({\bf k},j)\right]^{-1}-t_{\bf k}\right\}^{-1}.
\end{equation}
Here the statistical averaging denoted by the angular brackets and the time dependencies of operators
\begin{equation*}
\bar{a}_{{\bf l}\sigma}(\tau)=\exp({\cal H}\tau)a^\dagger_{{\bf l}\sigma}\exp(-{\cal H}\tau)
\end{equation*}
are determined by the operator ${\cal H}=H-\mu\sum_{{\bf l}\sigma}n_{{\bf l}\sigma}$ with the chemical potential $\mu$, ${\bf k}$ is the 3D wave vector, $j$ is an integer defining the fermionic Matsubara frequency $\omega_j=(2j-1)\pi T$ with the temperature $T$, $t_{\bf k}$ is the Fourier transform of $t_{\bf l'l}$, and ${\cal T}$ is the chronological operator.

\begin{figure}[t]
\centerline{\resizebox{0.8\columnwidth}{!}{\includegraphics{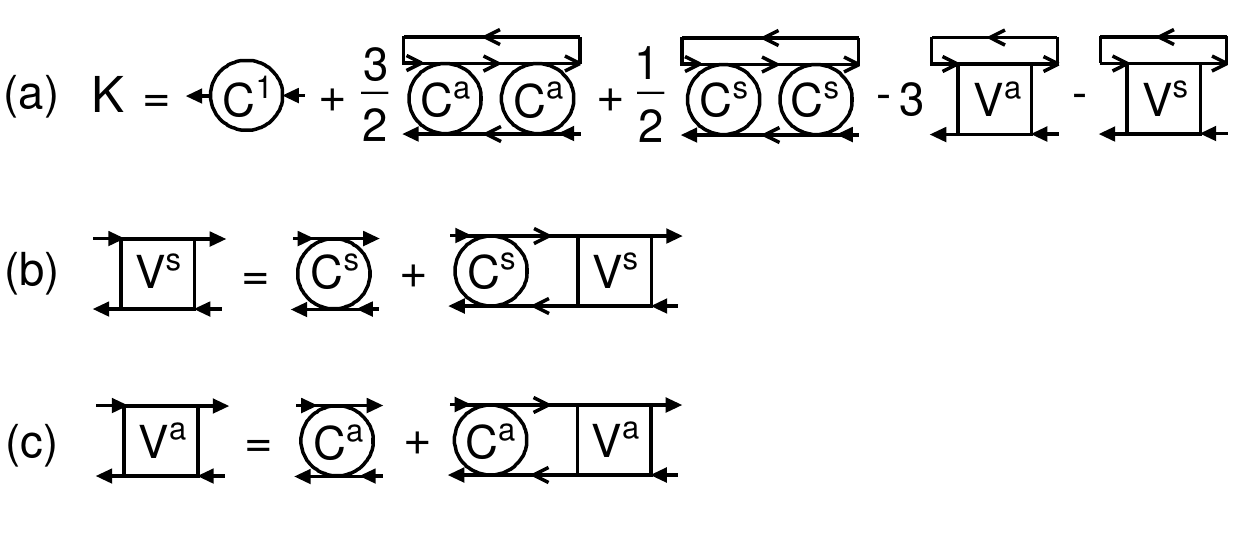}}}
\caption{(a) Diagrams taken into account in calculations of the irreducible part $K({\bf k},j)$. The circle with the notation $C^1$ is the first-order cumulant, circles with notations $C^s$ and $C^a$ are symmetrized and antisymmetrized second-order cumulants, lines with open arrows are renormalized hopping $\theta({\bf k},j)$,  squares with notations $V^s$ and $V^a$ are infinite sums of ladder diagrams symmetrized and antisymmetrized over spin indices. The Bethe-Salpeter equations that they satisfy are depicted in parts (b) and (c).} \label{Fig1}
\end{figure}
Diagrams taken into account in the present calculations are shown in Fig.~\ref{Fig1}. Here circles with notations $C^1$, $C^s$ and $C^a$ are the first- and second-order on-site cumulants \cite{Kubo},
\begin{eqnarray*}
&&C^{(1)}(\tau',\tau)=\big\langle{\cal T}\bar{a}_{{\bf l}\sigma}(\tau')a_{{\bf l}\sigma}(\tau)\big\rangle_0,\\
&&C^{(2)}(\tau_1,\sigma_1;\tau_2,\sigma_2;\tau_3,\sigma_3;\tau_4,\sigma_4)\\
&&\quad=\big\langle{\cal T}\bar{a}_{{\bf l}\sigma_1}(\tau_1)a_{{\bf l}\sigma_2}(\tau_2) \bar{a}_{{\bf l}\sigma_3}(\tau_3)a_{{\bf l}\sigma_4}(\tau_4)\big\rangle_0\\
&&\quad\quad-\big\langle{\cal T}\bar{a}_{{\bf l}\sigma_1}(\tau_1)a_{{\bf l}\sigma_2}(\tau_2)\big\rangle_0\big\langle{\cal T}\bar{a}_{{\bf l}\sigma_3}(\tau_3)a_{{\bf l}\sigma_4}(\tau_4)\big\rangle_0\\
&&\quad\quad+\big\langle{\cal T}\bar{a}_{{\bf l}\sigma_1}(\tau_1)a_{{\bf l}\sigma_4}(\tau_4)\big\rangle_0\big\langle{\cal T}\bar{a}_{{\bf l}\sigma_3}(\tau_3)a_{{\bf l}\sigma_2}(\tau_2)\big\rangle_0.
\end{eqnarray*}
The subscript 0 at angle brackets indicates that operator time dependencies and averaging are determined by the site Hamiltonian
\begin{equation*}
{\cal H}_{\bf l}=\sum_\sigma\big[(U/2)n_{\bf l\sigma}n_{\bf l,-\sigma}-\mu n_{\bf l\sigma}\big].
\end{equation*}
Superscripts $(s)$ and $(a)$ indicate symmetrization and antisymmetrization over spin indices of Fourier-transformed second-order cumulants,
\begin{eqnarray*}
&&C^{(s)}(j+\nu,j,j',j'+\nu)=\sum_{\sigma'}C^{(2)}(j+\nu,\sigma';j,\sigma;j',\sigma; j'+\nu,\sigma'),\\
&&C^{(a)}(j+\nu,j,j',j'+\nu)=\sum_{\sigma'}\sigma\sigma'C^{(2)}(j+\nu,\sigma';j, \sigma; j',\sigma;j'+\nu,\sigma'),
\end{eqnarray*}
with $\nu$ an integer defining the bosonic Matsubara frequency $\omega_{\nu}=2\nu\pi T$. In Fig.~\ref{Fig1}, block arrows entering and leaving cumulants and vertices shown by squares are their endpoints, lines with open arrows connecting these endpoints are the renormalized hopping,
\begin{equation}\label{theta}
\theta({\bf k},j)=t_{\bf k}+t_{\bf k}^2G({\bf k},j),
\end{equation}
vertices $V^{(s)}$ and $V^{(a)}$ are sums of infinite sequences of ladder diagrams, Fig.~\ref{Fig1}(b) and (c). In this work, only diagrams containing first- and second-order cumulants are considered.

The algebraic form of the graphical equations in Fig.~\ref{Fig1} is the following:
\begin{eqnarray}\label{K}
K({\bf k},j)&=&C^{(1)}(j)+\frac{T^2}{4N}\sum_{{\bf k'}j'\nu}\theta({\bf k'},j') {\cal T}_{\bf k-k'}(j+\nu,j'+\nu)\nonumber\\
&&\times\big[3C^{(a)}(j,j+\nu,j'+\nu,j')C^{(a)}(j+\nu,j,j',j'+\nu)\nonumber\\
&&\quad+C^{(s)}(j,j+\nu,j'+\nu,j')C^{(s)}(j+\nu,j,j',j'+\nu)\big]\nonumber\\
&&-\frac{T}{2N}\sum_{{\bf k'}j'}\theta({\bf k'},j')\big[3V_{\bf k-k'}^{(a)}(j,j,j',j')
+V_{\bf k-k'}^{(s)}(j,j,j',j')\big],
\end{eqnarray}
\begin{eqnarray}
&&V^{(i)}_{\bf k}(j+\nu,j,j',j'+\nu)=C^{(i)}(j+\nu,j,j',j'+\nu)\nonumber\\
&&\quad+T\sum_{\nu'} C^{(i)}(j+\nu,j+\nu',j'+\nu',j'+\nu){\cal T}_{\bf k}(j+\nu',j'+\nu') \nonumber\\
&&\quad\quad\times V^{(i)}_{\bf k}(j+\nu',j,j',j'+\nu'),\label{BSE}
\end{eqnarray}
where the superscript $i=s$ or $a$. ${\cal T}_{\bf k}(j,j')=N^{-1}\sum_{\bf k'}\theta({\bf k+k'},j)\theta({\bf k'},j')$, and $N$ is the number of sites.

Expressions for the first- and second-order cumulants can be found in Refs.~\cite{Vladimir,Metzner,Pairault,Sherman18}. In the case
\begin{equation}\label{conditions}
T\ll\mu,\quad T\ll U-\mu,
\end{equation}
they are significantly simplified. This range of chemical potentials includes half-filling, $\mu=U/2$, and moderate doping for $U\gg T$. In this range, cumulants read
\begin{eqnarray}\label{cumulants}
&&C^{(1)}(j)=\frac{1}{2}\big[g_1(j)+g_2(j)\big],\nonumber \\
&&C^{(2)}(j+\nu,\sigma;j,\sigma';j',\sigma';j'+\nu,\sigma)=\frac{1}{4T}\big[\delta_{jj'}\big(1-2 \delta_{\sigma\sigma'}\big)\nonumber\\
&&\quad+\delta_{\nu0}\big(2-\delta_{\sigma\sigma'}\big)\big]a_1(j'+\nu)a_1(j)-\frac{1}{2} \delta_{\sigma,-\sigma'}\big[a_1(j'+\nu)a_2(j,j')\\
&&\quad+a_2(j'+\nu,j+\nu)a_1(j)+a_3(j'+\nu,j+\nu)a_4(j,j')\nonumber\\
&&\quad+a_4(j'+\nu,j+\nu)a_3(j,j')\big],\nonumber
\end{eqnarray}
where
\begin{eqnarray*}
&&g_1(j)=({\rm i}\omega_j+\mu)^{-1},\quad g_2(j)=({\rm i}\omega_j+\mu-U)^{-1}, \\
&&a_1(j)=g_1(j)-g_2(j),\quad a_2(j,j')=g_1(j)g_1(j'),\\
&&a_3(j,j')=g_2(j)-g_1(j'),\quad a_4(j,j')=a_1(j)g_2(j').
\end{eqnarray*}

With expressions~(\ref{cumulants}), vertices $V^{(s)}$ and $V^{(a)}$ acquire the form
\begin{eqnarray}
&&V_{\bf k}^{(s)}(j+\nu,j,j',j'+\nu)=\frac{1}{2}f_{\bf k}^{(2)}(j+\nu,j'+\nu) \nonumber\\
&&\quad\times\big\{2C^{(s)}(j+\nu,j,j',j'+\nu)-a_2(j'+\nu,j+\nu)z_1({\bf k},j,j')\nonumber\\
&&\quad-a_1(j'+\nu)z_2({\bf k},j,j')-a_4(j'+\nu,j+\nu)z_3({\bf k},j,j')\nonumber\\
&&\quad-a_3(j'+\nu,j+\nu)z_4({\bf k},j,j')\big\},\label{Vs}\\
&&V_{\bf k}^{(a)}(j+\nu,j,j',j'+\nu)=\frac{1}{2}f_{\bf k}^{(1)}(j+\nu,j'+\nu) \nonumber\\
&&\quad\times\big\{2C^{(a)}(j+\nu,j,j',j'+\nu)+\big[a_2(j'+\nu,j+\nu)\nonumber\\
&&\quad-\frac{1}{T}\delta_{jj'}a_1(j'+\nu)\big]y_1({\bf k},j,j')+a_1(j'+\nu)y_2({\bf k},j,j') \nonumber\\
&&\quad+a_4(j'+\nu,j+\nu)y_3({\bf k},j,j')+a_3(j'+\nu,j+\nu)y_4({\bf k},j,j')\big\},\label{Va}
\end{eqnarray}
where
\begin{eqnarray*}
&&f^{(1)}_{\bf k}(j,j')=\bigg[1+\frac{1}{4}a_1(j)a_1(j'){\cal T}_{\bf k}(j,j')\bigg]^{-1},\\
&&f^{(2)}_{\bf k}(j,j')=\bigg[1-\frac{3}{4}a_1(j)a_1(j'){\cal T}_{\bf k}(j,j')\bigg]^{-1},
\end{eqnarray*}
and quantities $z_i$ and $y_i$, $i=1,\ldots 4$ satisfy two systems of linear equations
\begin{eqnarray}
&&z_i({\bf k},j,j')=d_i({\bf k},j,j')-e_{i2}({\bf k},j-j')z_1({\bf k},j,j')\nonumber \\
&&\quad-e_{i1}({\bf k},j-j')z_2({\bf k},j,j')-e_{i4}({\bf k},j-j')z_3({\bf k},j,j')\nonumber\\
&&\quad-e_{i3}({\bf k},j-j')z_4({\bf k},j,j'),\label{zi}\\
&&y_i({\bf k},j,j')=b_i({\bf k},j,j')+\big[c_{i2}({\bf k},j-j')-T^{-1}\delta_{jj'}c_{i1}({\bf k},j-j')\big]\nonumber \\
&&\quad\times y_1({\bf k},j,j')+c_{i1}({\bf k},j-j')y_2({\bf k},j,j')+c_{i4}({\bf k},j-j')y_3({\bf k},j,j')\nonumber\\
&&\quad+c_{i3}({\bf k},j-j')y_4({\bf k},j,j')\label{yi}
\end{eqnarray}
with the coefficients
\begin{eqnarray*}
&&e_{ii'}({\bf k},\nu)=\frac{T}{2}\sum_j a_i(\nu+j,j)a_{i'}(j,\nu+j){\cal T}_{\bf k}(\nu+j,j) f^{(2)}_{\bf k}(\nu+j,j),\\
&&c_{ii'}({\bf k},\nu)=\frac{T}{2}\sum_j a_i(\nu+j,j)a_{i'}(j,\nu+j){\cal T}_{\bf k}(\nu+j,j) f^{(1)}_{\bf k}(\nu+j,j),\\
&&d_i({\bf k},j,j')=\frac{3}{4}a_i(j,j')a_1(j)a_1(j'){\cal T}_{\bf k}(j,j')f^{(2)}_{\bf k}(j,j')\\ &&\quad-e_{i1}({\bf k},j-j')a_2(j,j')-e_{i2}({\bf k},j-j')a_1(j)\\
&&\quad-e_{i3}({\bf k},j-j')a_4(j,j')-e_{i4}({\bf k},j-j')a_3(j,j'),\\
&&b_i({\bf k},j,j')=-\frac{1}{4}a_i(j,j')a_1(j)a_1(j'){\cal T}_{\bf k}(j,j')f^{(1)}_{\bf k}(j,j')\\ &&\quad+c_{i1}({\bf k},j-j')\big[a_2(j,j')-T^{-1}\delta_{jj'}a_1(j)\big]+c_{i2}({\bf k},j-j')a_1(j)\\
&&\quad+c_{i3}({\bf k},j-j')a_4(j,j')+c_{i4}({\bf k},j-j')a_3(j,j').
\end{eqnarray*}
Hence the two BSE's (\ref{BSE}) for $V^{(s)}$ and $V^{(a)}$ are reduced to two small systems of linear equations (\ref{zi}) and (\ref{yi}) -- each system contains four equations for four variables for given values of ${\bf k}$, $j$ and $j'$. They can be solved exactly.

If the main determinant $\Delta({\bf k},\nu)$ of either system of linear equations vanishes for some values of ${\bf k}$ and $\nu=j-j'$ (note that the main determinants depend only on these two variables), this will lead to the divergence of the respective vertex (\ref{Vs}) or (\ref{Va}), and thus to the divergence of the spin or charge susceptibilities
\begin{eqnarray}
\chi^{\rm sp}({\bf k},\nu)&=&\int_{0}^{\beta}{\rm e}^{-{\rm i}\omega_\nu \tau}\sum_{\bf l}{\rm e}^{{\rm i}{\bf kl}}\langle{\cal T}s^\sigma_{\bf l}(\tau)s^{-\sigma}_{\bf 0}\rangle{\rm d}\tau\nonumber\\
&=&-\frac{T}{N}\sum_{{\bf k'}j}G({\bf k+k'},\nu+j)G({\bf k'},j)-T^2\sum_{jj'}F_{\bf k}(j,\nu+j) \nonumber\\
&&\times F_{\bf k}(j',\nu+j')V^{(a)}_{\bf k}(j+\nu,j'+\nu,j',j),\label{chi_sp} \\
\chi^{\rm ch}({\bf k},\nu)&=&\int_{0}^{\beta}{\rm e}^{-{\rm i}\omega_\nu \tau}\sum_{\bf l}{\rm e}^{{\rm i}{\bf kl}}\frac{1}{2}\langle{\cal T}(n_{\bf l}(\tau)-\bar{n})(n_{\bf 0}-\bar{n})\rangle{\rm d}\tau\nonumber\\
&=&-\frac{T}{N}\sum_{{\bf k'}j}G({\bf k+k'},\nu+j)G({\bf k'},j)-T^2\sum_{jj'}F_{\bf k}(j,\nu+j) \nonumber\\
&&\times F_{\bf k}(j',\nu+j')V^{(s)}_{\bf k}(j+\nu,j'+\nu,j',j).\label{chi_ch}
\end{eqnarray}
Here $s^\sigma_{\bf l}=a^\dagger_{{\bf l}\sigma}a_{{\bf l},-\sigma}$, $n_{\bf l}=\sum_\sigma n_{{\bf l}\sigma}$, $\bar{n}=\langle n_{\bf l}\rangle$, $\beta=1/T$,
\begin{equation*}
 F_{\bf k}(j,j')=N^{-1}\sum_{\bf k'}\Pi({\bf k'},j)\Pi({\bf k+k'},j'),\,{\rm and}\;\Pi({\bf k},j)=1+t_{\bf k}G({\bf k},j).
\end{equation*}
Thus, the vanishing $\Delta({\bf k},0)$ in one of the two BSE's (\ref{zi}) or (\ref{yi}) indicates a phase transition in charge or spin subsystem, and the wave vector at which it happens determines the symmetry of the ordered phase. As follows from the above formulas, the determinants are dimensionless and real quantities. In the considered ranges of parameters the determinant of the system (\ref{zi}), which describes charge excitations, is of the order of unity for all {\bf k}, while the determinant in (\ref{yi}) related to the spin subsystem becomes small at some $T$ for all considered values of $U$ and $\bar{n}$ at the momentum ${\bf k=Q}=(\pi,\pi,\pi)$ (the intersite distance is set as the unit of length). That is, the ordered state is antiferromagnetic.

We emphasize that the equations of this section are the same as those used for investigating the 2D Hubbard model \cite{Sherman18,Sherman21,Sherman24}. The only difference is in the dimensionality of the wave vector {\bf k}.

\section{Results and Discussion}
The set of equations in the previous section was solved by iteration. As starting value for $G({\bf k},j)$, we used either the Green's function obtained from Eq.~(\ref{Larkin}) with $K({\bf k},j)=C^{(1)}(j)$, the first term on the right-hand side of Eq.~(\ref{K}), or a converged value obtained for a somewhat larger $T$. We considered the Hubbard repulsion range $4t\leq U\leq12t$ in 8$^3$ and 10$^3$ lattices. The ranges of $T$ and $\bar{n}$ were limited by conditions~(\ref{conditions}).

\subsection{$\Delta({\bf Q},0)$ near Zero}
\begin{figure}[t]
\centerline{\resizebox{0.8\columnwidth}{!}{\includegraphics{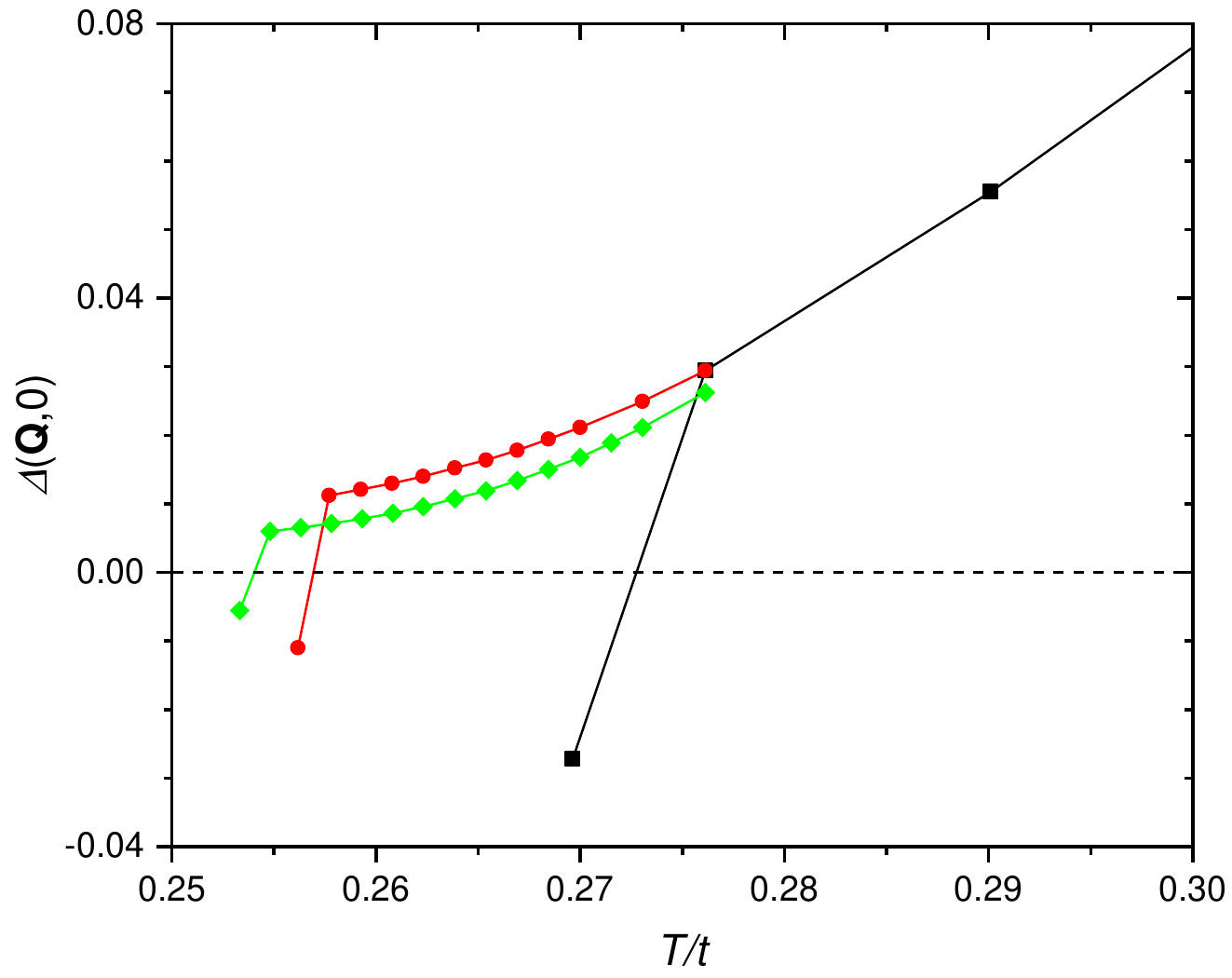}}}
\caption{The temperature dependence of the main determinant $\Delta({\bf Q},0)$ of the system of linear equations (\ref{yi}) for $U=4t$. Data shown by black squares were obtained with the iteration starting from $K({\bf k},j)=C^{(1)}(j)$ in a 8$^3$ lattice, red circles are calculated in the same lattice with the starting $G({\bf k},j)$ taken from a result with a slightly larger $T$, green rhombi are data analogously obtained in a 10$^3$ lattice.}\label{Fig2}
\end{figure}
From textbooks, it is known that transitions between the para- and ferro- or antiferromagnetic phases are second-order. It is true for an infinite crystal. Let us consider the static magnetic susceptibility at the antiferromagnetic wave vector {\bf Q},
\begin{equation}\label{static}
\chi^{\rm sp}({\bf Q},0)=\sum_{\bf l}{\rm e}^{{\rm i}{\bf Ql}}\int_{0}^{\beta}\langle{\cal T}s^\sigma_{\bf l}(\tau)s^{-\sigma}_{\bf 0}\rangle{\rm d}\tau.
\end{equation}
Each term in the above lattice sum is finite, and if the number of sites in a sample is finite, the susceptibility remains finite for all $T$, reaching a maximum value for $T=T_{\rm N}$. An infinite value of the sum (\ref{static}) can be obtained only in an infinite crystal. In a finite lattice, a further decrease of $T$ below $T_{\rm N}$ will lead to an abrupt sign change in the susceptibility. For the quantity $[\chi^{\rm sp}({\bf Q},0)]^{-1}$ usually used for determining $T_{\rm N}$ this fact reveals itself in a small jump across zero. It is shown in Fig.~\ref{Fig2} where the temperature dependence of the main determinant $\Delta({\bf Q},0)$ of the magnetic BSE (\ref{yi}) is depicted. From the above formulas, we know that $[\chi^{\rm sp}({\bf Q},0)]^{-1}\sim\Delta({\bf Q},0)$. Notice that the spin susceptibility (\ref{chi_sp}) is defined such that $\chi^{\rm sp}({\bf k},\nu)$ is a real, nonnegative, and even function of $\nu$. Hence, the value below zero in Fig.~\ref{Fig2} is nonphysical. It only indicates the transition to an ordered state, which cannot be described by the above formulas because they belong to the paramagnetic state. However, the fact that a smooth approach of $[\chi^{\rm sp}({\bf Q},0)]^{-1}$ to zero is transformed to a small jump shows that the second-order phase transition has some features of the first-order transition in a finite crystal. As follows from Fig.~\ref{Fig2}, the jump magnitude decreases with growing the sample size, and the jump ceases to be detectable when its magnitude becomes smaller than the resolution of a calculation method. Apparently, the difference in the conclusion about the weak first-order character of the transitions between Refs.~\cite{Hirsch,Scalettar} and \cite{Staudt} is connected with different methods of determining $T_{\rm N}$ and sample sizes.

\subsection{$T_{\rm N}$ at Half-filling}
\begin{figure}[t]
\centerline{\resizebox{0.8\columnwidth}{!}{\includegraphics{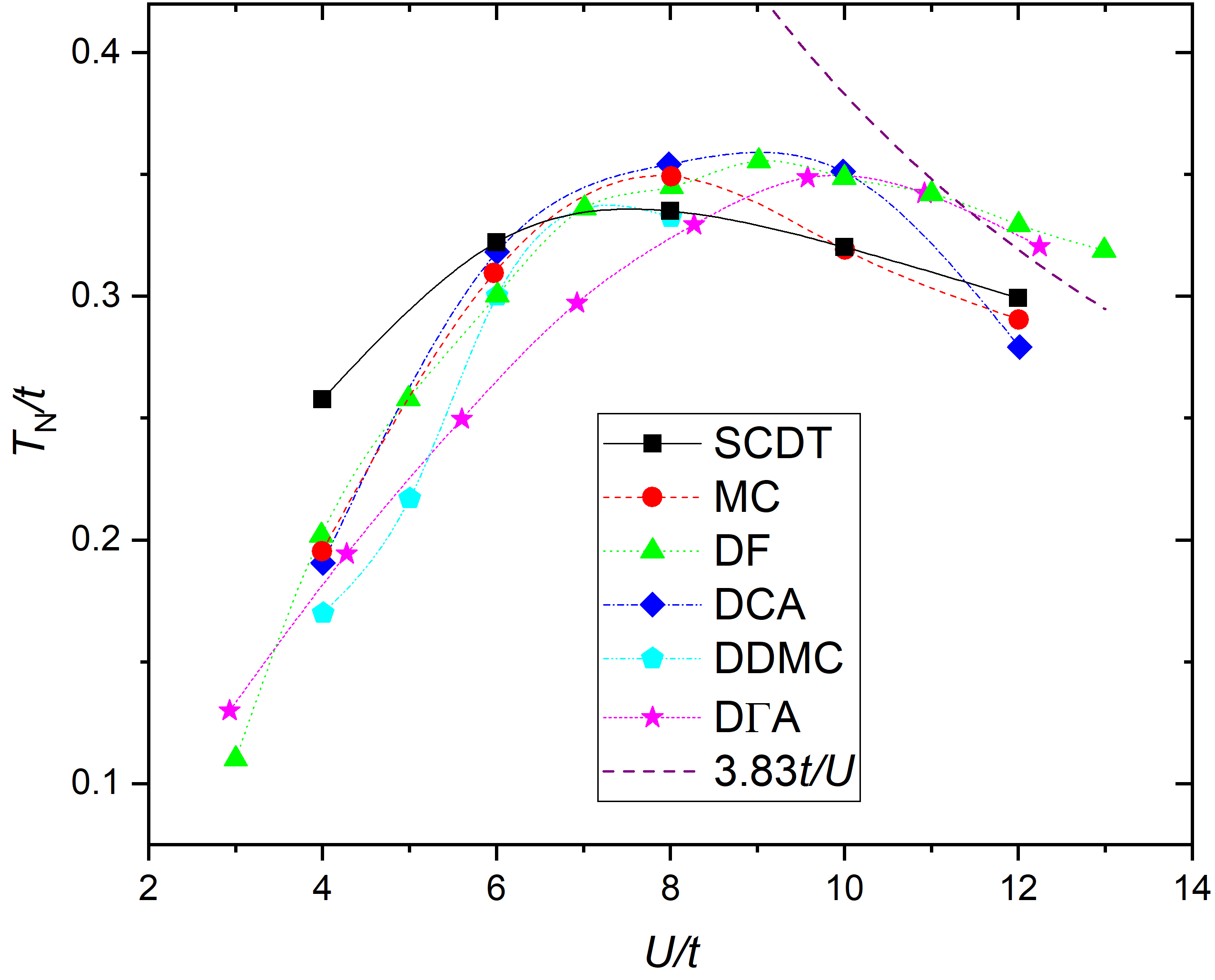}}}
\caption{The dependence of the N\'eel temperature $T_{\rm N}$ on the Hubbard repulsion $U$ calculated with SCDT in a half-filled 8$^3$ lattice is shown by black squares. The comparison data are obtained by quantum Monte Carlo simulations \cite{Staudt} (red circles), the dual fermion approach \cite{Hirschmeier} (green triangles), dynamic cluster approximation \cite{Kent} (blue rhombi), determinantal diagrammatic Monte Carlo \cite{Kozik} (cyan pentagons), and dynamical vertex approximation \cite{Rohringer} (magenta stars). The purple dashed line shows the Heisenberg limit $T_{\rm N}=3.83t^2/U$ \cite{Rushbrooke,Sandvik}.}\label{Fig3}
\end{figure}

The dependence $T_{\rm N}(U)$ calculated with SCDT in a half-filled 8$^3$ lattice is shown in Fig.~\ref{Fig3}. For comparison, results of several other approaches are also depicted together with this dependence in the spin-$\frac{1}{2}$ Heisenberg model \cite{Rushbrooke,Sandvik}. As seen from the figure, for $U\geq6t$, our results are close to the majority of other data and tend to the Heisenberg limit for large repulsions. This could be expected, since SCDT is designed for strong interaction. On the other hand, we see that our result for $U=4t$ noticeably exceeds other data. The first and main reason for the difference is the fact that this case belongs to weak coupling, which is not well described by SCDT. The second reason is connected with the fact that $T_{\rm N}$ in Fig.~\ref{Fig3} belongs to finite clusters. It corresponds to the lowest achieved temperature, for which $[\chi^{\rm sp}({\bf Q},0)]^{-1}$ remains positive. As the magnetic correlation length $\xi$ becomes comparable to the cluster size, this quantity jumps to a negative value. Therefore, $T_{\rm N}$ in Fig.~\ref{Fig3} is somewhat larger than the N\'eel temperature in an infinite crystal. The correlation length can be roughly estimated using the formula \cite{Hasenfratz} for the Heisenberg model $\xi\sim\rho_s^{-1}\exp(2\pi\rho_s/T)$, where the spin stiffness $\rho_s\sim J\sim U^{-1}$ with the exchange constant $J=4t^2/U$. Hence, for low $T$, the correlation length is larger for a smaller $U$ and the same other parameters. Therefore, for a smaller $U$, the increase of $T_{\rm N}$ due to the lattice size reduction is more pronounced than for larger repulsions. This fact can contribute to the mentioned difference of our obtained $T_{\rm N}$ with the results of other approaches.

\begin{figure}[t]
\centerline{\resizebox{0.6\columnwidth}{!}{\includegraphics{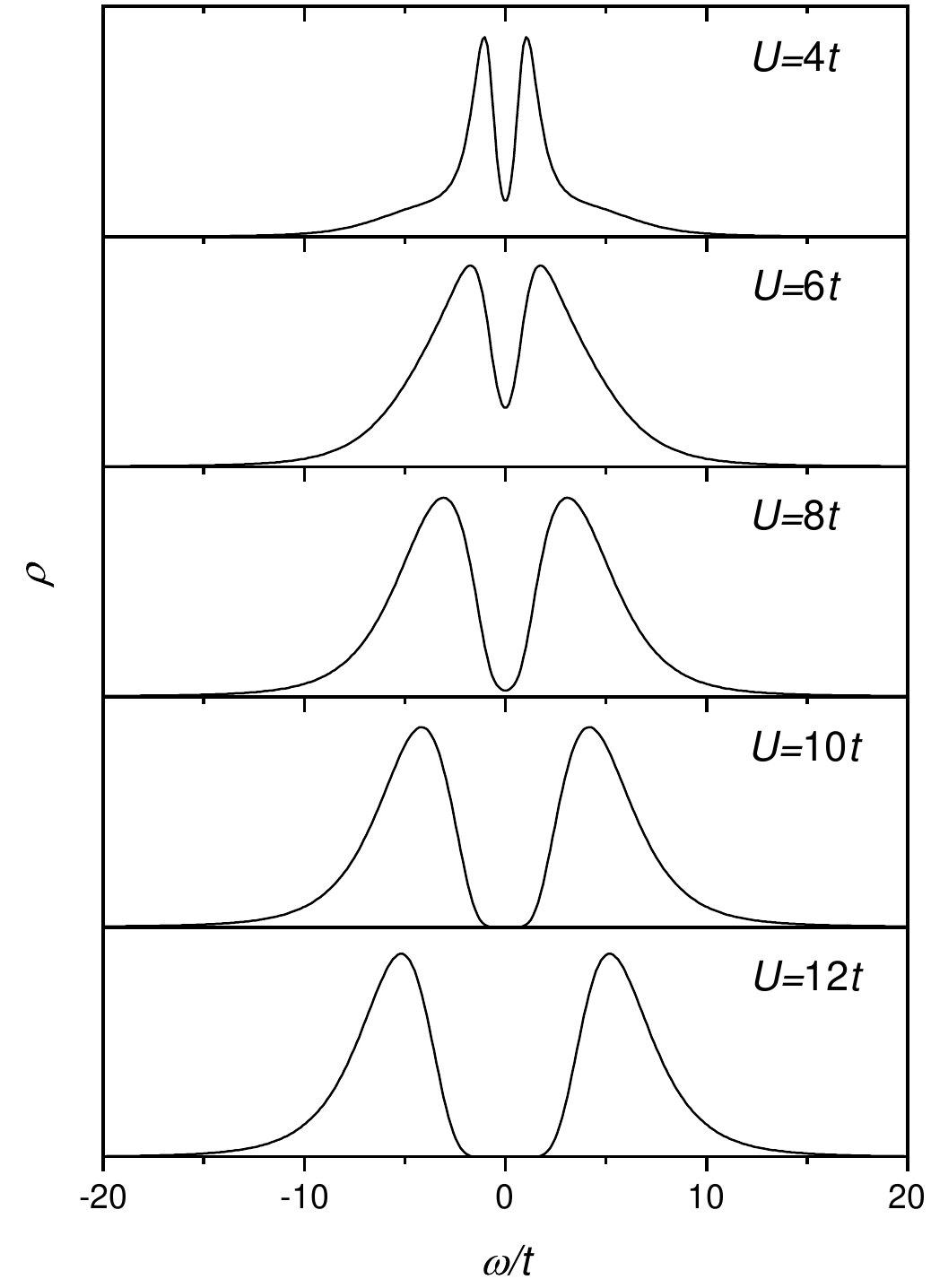}}}
\caption{The density of states for the repulsion values indicated in the panels and the respective $T=T_{\rm N}$ shown in Fig.~\ref{Fig3}.}\label{Fig4}
\end{figure}
Figure~\ref{Fig4} demonstrates densities of states (DOS),
\begin{equation*}
\rho(\omega)=-\frac{1}{\pi N}\sum_{\bf k}{\rm Im}\,G({\bf k},\omega),
\end{equation*}
calculated for several values of the Hubbard repulsion and the respective temperatures $T=T_{\rm N}$ shown in Fig.~\ref{Fig3}. The analytic continuation of Green's functions from imaginary to real frequencies $\omega$ was performed using the maximum entropy method \cite{Press,Jarrell96,Habershon}. From Fig.~\ref{Fig4}, one can see that the DOS at $U=4t$ differs essentially from spectra for larger repulsions. In the former case, the DOS is well separated into an intensive quasiparticle maximum with a dip near $\omega=0$ originating in part from the Hubbard repulsion and the Slater mechanism \cite{Slater} and a less intensive higher-frequency electron-spin-excitation continuum. Such a type of DOS is inherent in weak coupling. For $U\geq6t$, it is difficult to separate the spectrum into these two parts -- the case is typical for strong correlations. With their growth, the dip near $\omega=0$ goes down, and for $U\approx9t$ it transforms into the Mott gap. Notice that we are dealing with the comparatively high temperature, $T_{\rm N}\approx0.3t$ for this $U$ (see Fig.~\ref{Fig3}). In the 2D case, for such $T$, the Mott gap appears at $U\approx7t$. In this case, for $T\sim0.01t$, the respective $U$ decreases to $5.5t$ where the Mott gap is smoothly transformed into the Slater gap \cite{Sherman23b}.

\subsection{Doping}
\begin{figure}[t]
\centerline{\resizebox{0.8\columnwidth}{!}{\includegraphics{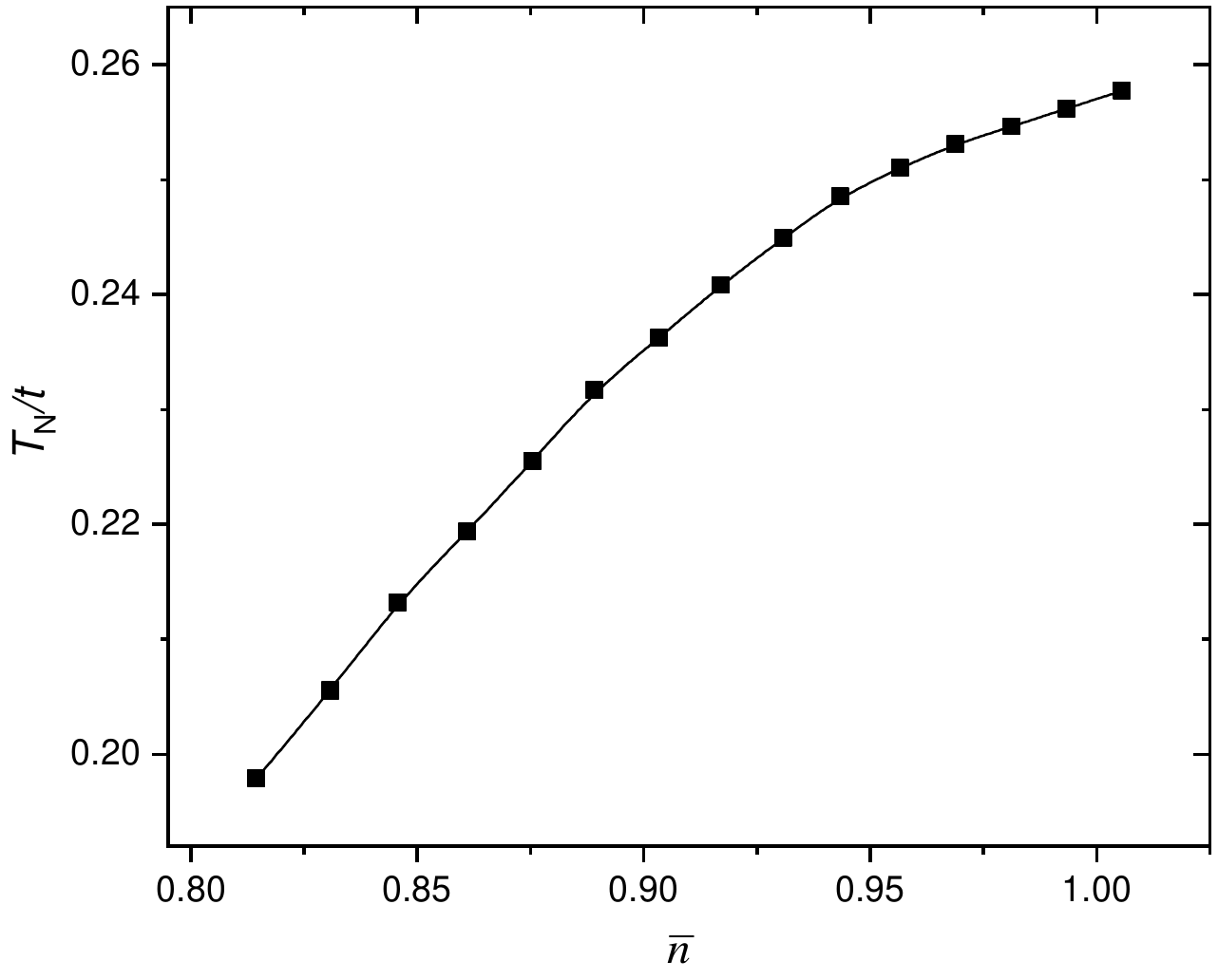}}}
\caption{The variation of $T_{\rm N}$ with doping in the case $U=4t$.}\label{Fig5}
\end{figure}
\begin{figure}[h]
\centerline{\resizebox{0.8\columnwidth}{!}{\includegraphics{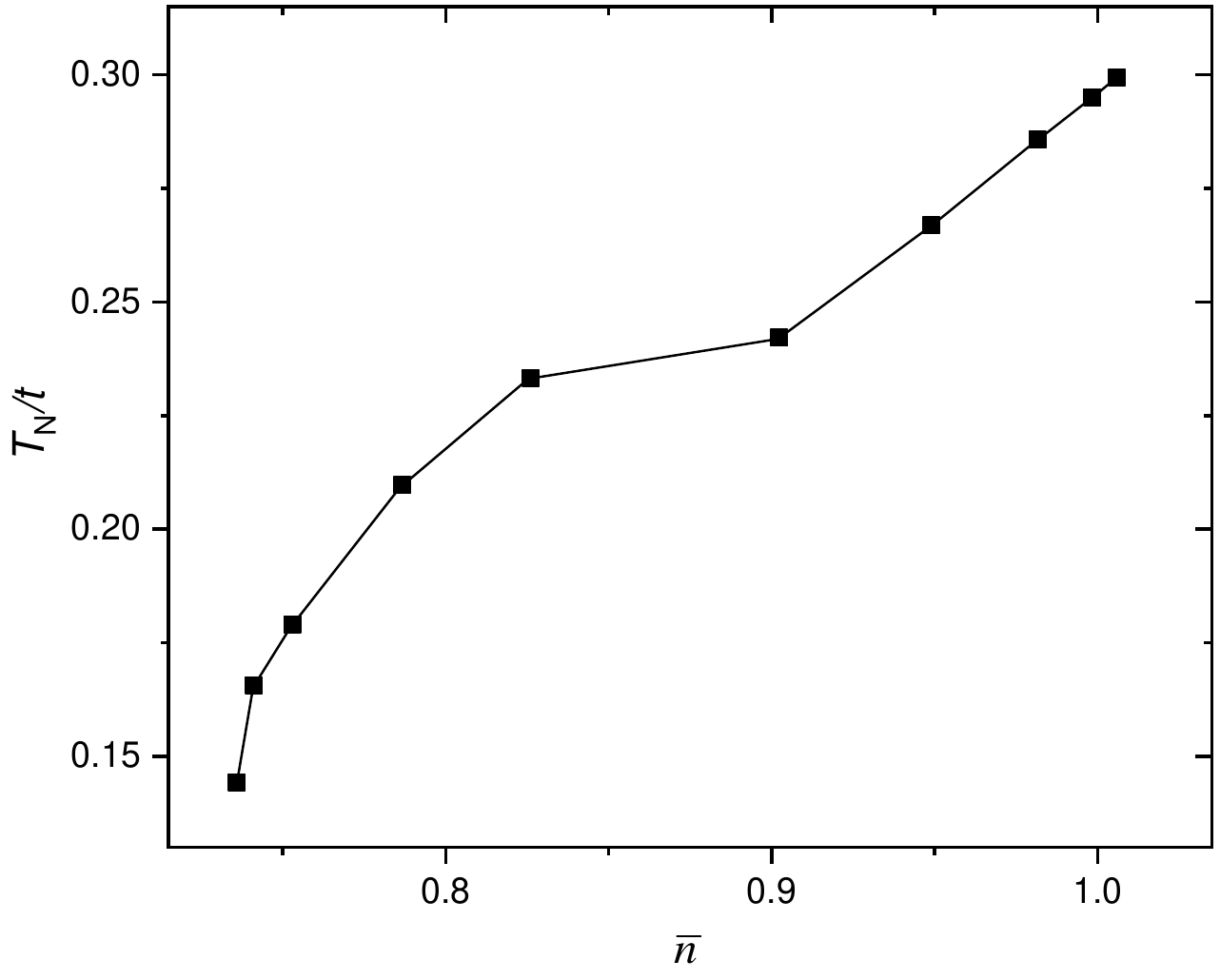}}}
\caption{The variation of $T_{\rm N}$ with doping in the case $U=12t$.}\label{Fig6}
\end{figure}
The change of the N\'eel temperature with doping is shown in Figs.~\ref{Fig5} and~\ref{Fig6} for the cases $U=4t$ and $12t$, respectively. Since the Hamiltonian (\ref{Hubbard}) possesses the particle-hole symmetry, only chemical potentials $\mu\leq U/2$ and electron concentrations $\bar{n}\leq1$ are considered. Doping ranges in these figures are restricted by the conditions (\ref{conditions}), for which expressions (\ref{cumulants}) for cumulants are valid. In these ranges, the phase transition occurs between the paramagnetic and antiferromagnetic states.

\begin{figure}[h]
\centerline{\resizebox{0.6\columnwidth}{!}{\includegraphics{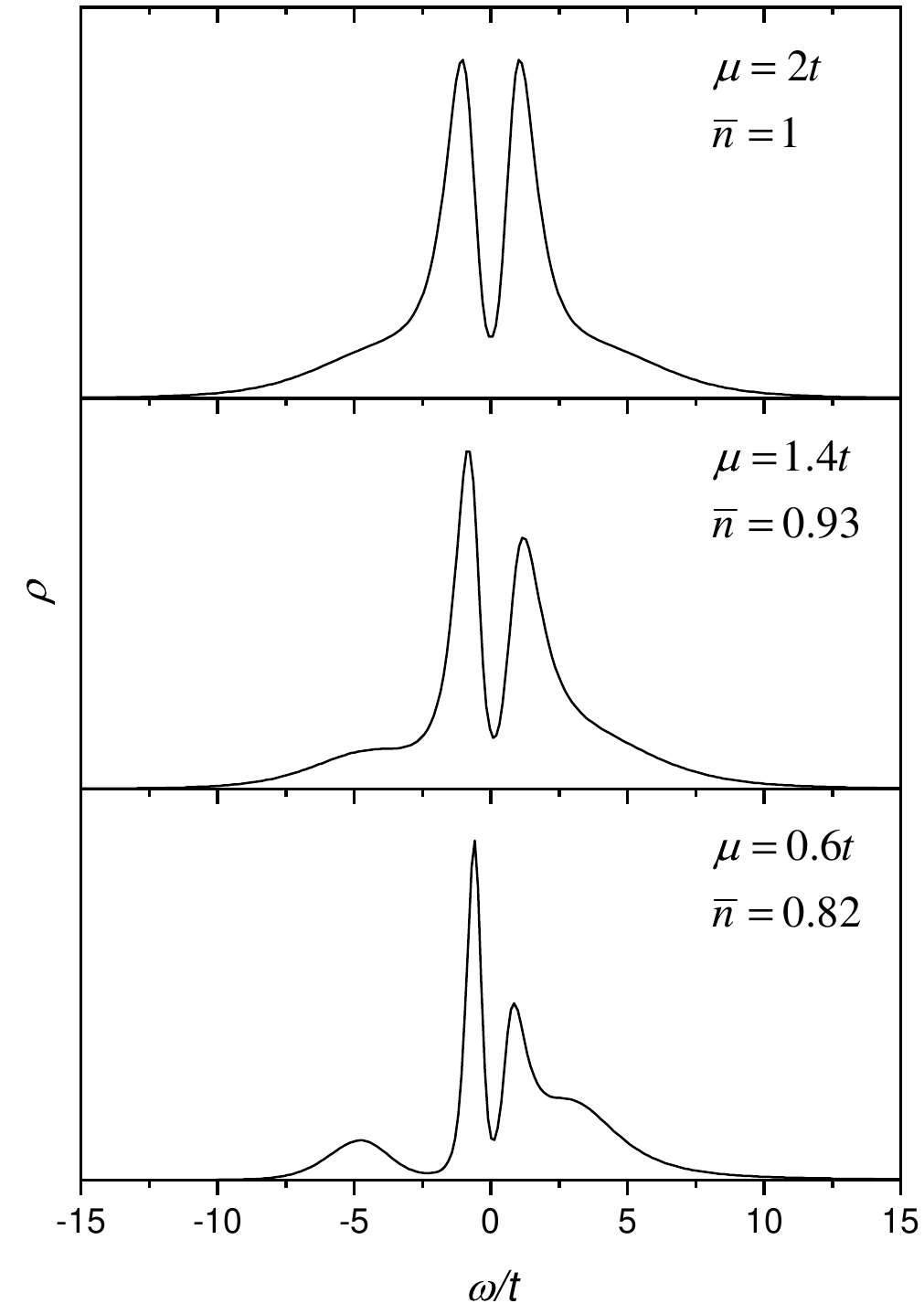}}}
\caption{The variation of the DOS with doping in the case $U=4t$. Values of the chemical potential and electron concentration are indicated on panels. Temperatures correspond to the phase-separation line in Fig.\ref{Fig5}.}\label{Fig7}
\end{figure}
\begin{figure}[h]
\centerline{\resizebox{0.6\columnwidth}{!}{\includegraphics{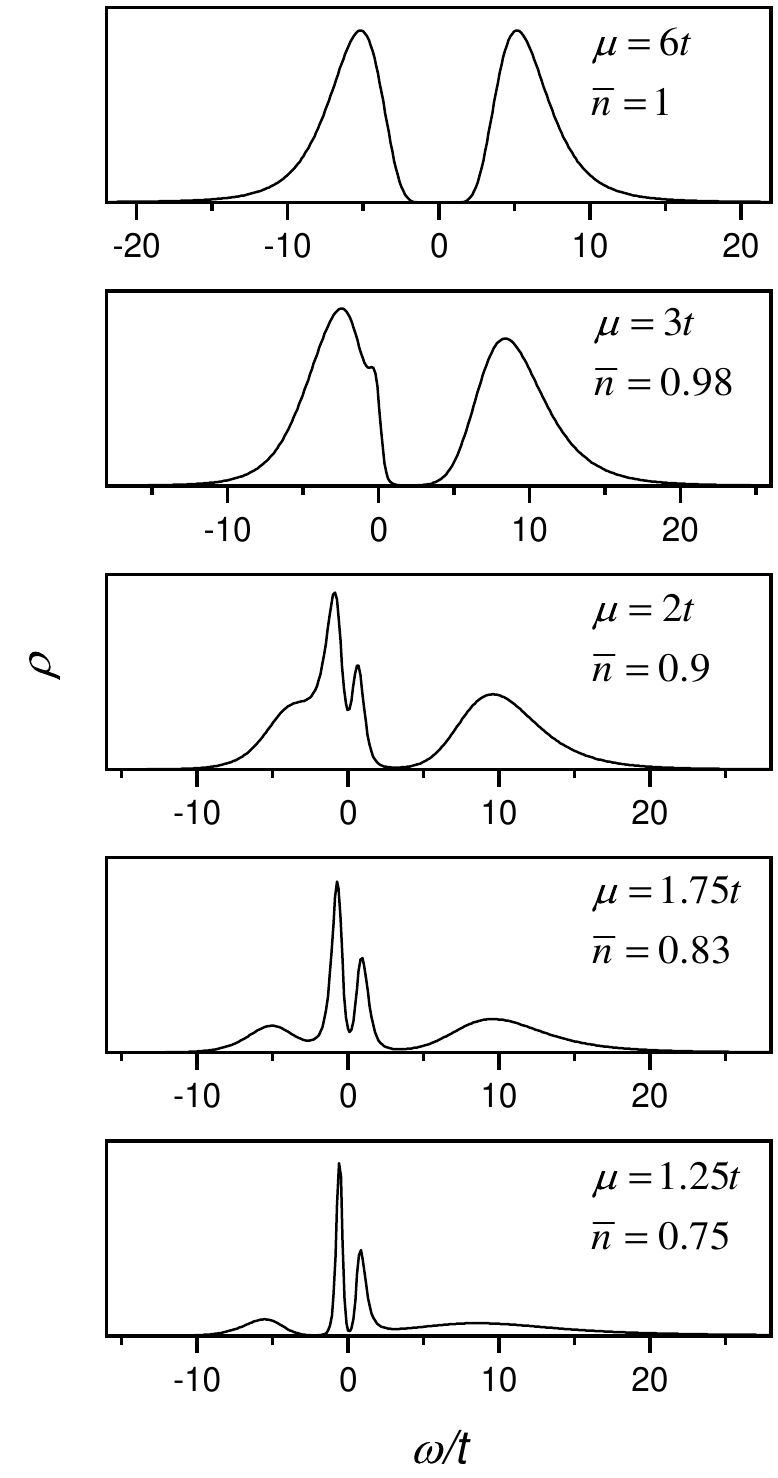}}}
\caption{The variation of the DOS with doping in the case $U=12t$. Values of the chemical potential and electron concentration are indicated on panels. Temperatures correspond to the phase-separation line in Fig.\ref{Fig6}.}\label{Fig8}
\end{figure}
For $U=4t$, Fig.~\ref{Fig5}, the dependence $T_{\rm N}(\bar{n})$ is monotonous, while for $U=12t$ in Fig.~\ref{Fig6} it demonstrates a plateau near $\bar{n}=0.87$. The reason of this difference can be understood from the evolution of DOSs with doping shown in Figs.~\ref{Fig7} and~\ref{Fig8}. The spectra demonstrated in these figures were calculated for parameters on the phase-separation lines in Figs.~\ref{Fig5} and~\ref{Fig6}. In the case $U=4t$ with doping, the spectrum character is unchanged. For all considered $\bar{n}$, the DOS has typical features of weak coupling consisting of an intensive quasiparticle maximum with a dip near $\omega=0$ and a less intensive higher frequency electron-spin-excitation continuum. Doping only redistributes the intensity between positive- and negative-frequency parts of the spectrum. For $U=12t$, the situation is different -- with doping, the character of the spectrum is changed. In the range $1\geq\bar{n}\gtrsim0.9$, it is a typical strong-coupling DOS with the Mott gap near its central part. However, for $\bar{n}\lesssim0.83$, the spectrum resembles spectra in Fig.~\ref{Fig7}, that is, the DOS acquires features typical of weak coupling. This fact can be interpreted as follows: the electron depopulation leads to rarefying the double occupancy of sites, which means an effective decay of the interaction. Hence, the plateau near $\bar{n}=0.87$ in Fig.~\ref{Fig6} is connected with the change in the DOS character from the strong- to weak-coupling type.

\subsection{Susceptibility Critical Exponent}
\begin{figure}[h]
\centerline{\resizebox{0.8\columnwidth}{!}{\includegraphics{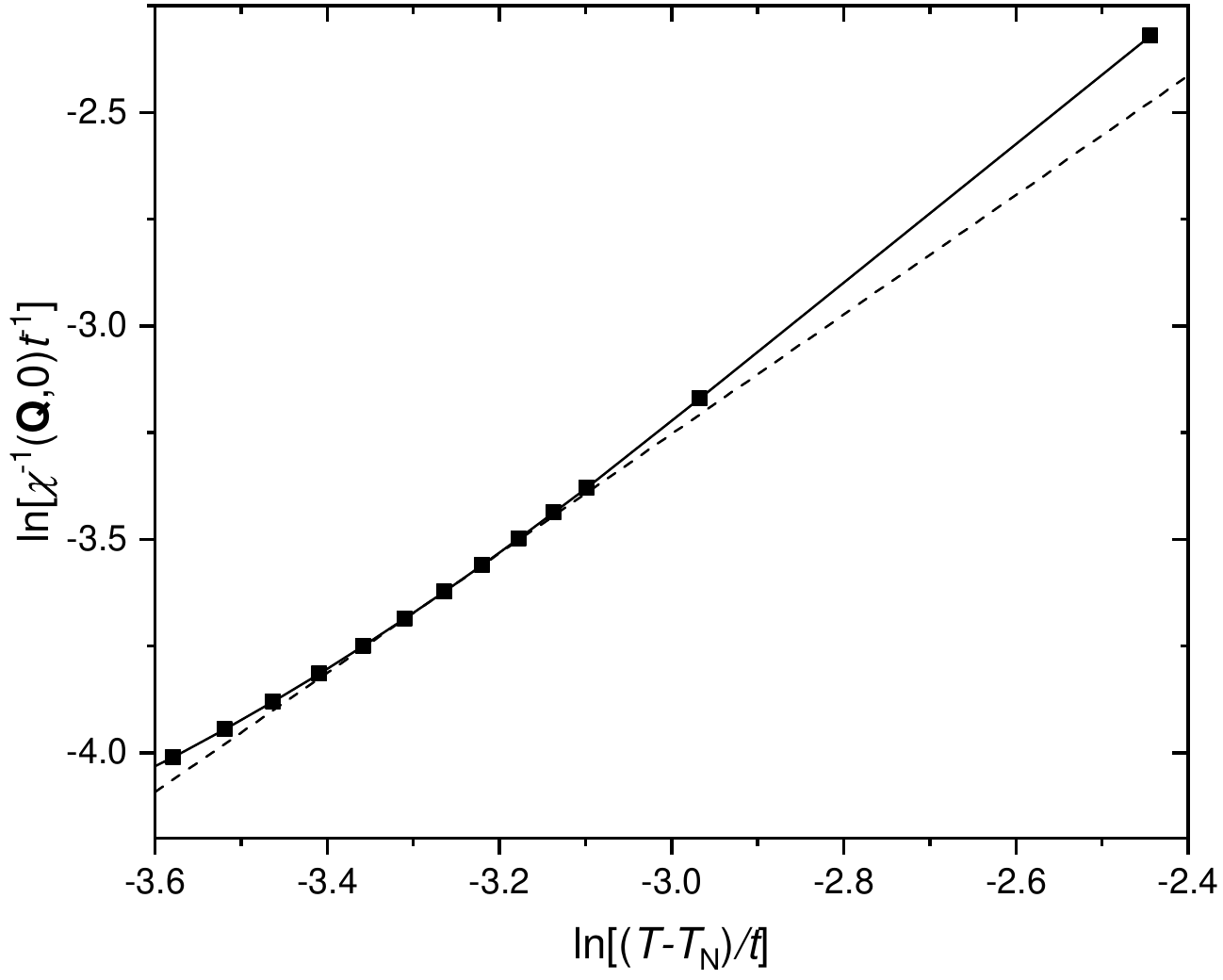}}}
\caption{The temperature dependence of the inverse spin susceptibility $[\chi^{\rm sp}({\bf Q},0)]^{-1}$ in the fluctuation region for half-filling, a 8$^3$ lattice, and $U=12t$ (the solid line and squares). The dashed line is the dependence $[\chi^{\rm sp}({\bf Q},0)]^{-1}=a(T-T_{\rm N})^\gamma$ with $\gamma=1.4$. $T_{\rm N}=0.299t$.}\label{Fig9}
\end{figure}
As seen from Fig.~\ref{Fig3}, for $U=12t$ and half-filling, the N\'eel temperature is close to that in the Heisenberg model with the related exchange constant. One can expect that the critical behavior of the susceptibility in fluctuation regions of the two models is also similar and satisfies the power-law dependence $\chi^{\rm sp}({\bf Q},0)\sim(T-T_{\rm N})^{-\gamma}$ with $\gamma\approx1.4$ \cite{Holm}. The calculated susceptibility and this power-law dependence are compared in Fig.~\ref{Fig9}. As can be seen, the two curves are indeed close in the higher-temperature part of the fluctuation region. However, they start to deviate at lower temperatures. The origin of this deviation is the fact, mentioned in the beginning of this section, that for a finite crystal $[\chi^{\rm sp}({\bf Q},0)]^{-1}\sim\Delta({\bf Q},0)$ does not vanish smoothly to zero as $T\rightarrow T_{\rm N}$, but rather jumps across zero. This behavior is shown in Fig.~\ref{Fig2}. A small but finite value of $[\chi^{\rm sp}({\bf Q},0)]^{-1}$ at the jump is the cause of the deviation in Fig.~\ref{Fig9}.

Critical exponents were analogously determined for other values of $U$ for half-filling. We found that $\gamma$ monotonously decreases as the repulsion is reduced. For $U=4t$, the critical exponent is close to unity. Similar results were obtained in Refs.~\cite{Rohringer,Hirschmeier}.

\section{Concluding Remarks}
In this work, we investigated the 3D fermionic Hubbard model on a simple cubic lattice in the case when the hopping integral is nonzero only between neighboring sites. For this purpose, the strong coupling diagram technique was used, in which infinite series of ladder diagrams were taken into account. In the range $4t\leq U\leq 12t$, this approach allows us to determine the phase boundary between paramagnetic and antiferromagnetic phases both for cases of half-filling and doping, to investigate the variation of the DOS on this boundary with changing the Hubbard repulsion and doping, and to calculate the magnetic critical exponent. For Hubbard repulsions $U\gtrsim6t$, our calculated N\'eel temperatures are close to those obtained by Monte Carlo simulations and some other methods. We found that along the phase boundary, the Mott transition occurs at $U\approx9t$. For $U=4t$, the N\'eel temperature monotonously decreases with doping. For $U=12t$, a plateau near the electron concentration $\bar{n}=0.87$ is observed in this dependence. As follows from DOS shapes, the plateau is connected with the spectrum reconstruction from the strong-coupling type with the Mott gap at small $|1-\bar{n}|$, to the weak-coupling types with an intensive central maximum and a weaker electron-spin-excitation continuum for larger doping. The reconstruction follows from rarifying the site double occupancy at large doping, which results in an effective reduction of the electron interaction. For $U=12t$ and half-filling, the obtained N\'eel temperature is close to $T_{\rm N}$ in the Heisenberg model with the respective exchange constant. We found that the magnetic critical exponents $\gamma$ of the two models are also close, being equal to $1.4$. As $U$ is decreased, $\gamma$ is monotonously reduced, tending unity for $U=4t$. We notice that for a finite crystal, the phase transition between paramagnetic and antiferromagnetic phases has some features of a first-order transition, which weaken as the crystal size grows.

\backmatter

\section*{Declarations}
Conflict of interest: The author declares no conflict of interest. There are no other applicable declarations.


\begin{thebibliography}{99}
\bibitem{Hirsch}J.E. Hirsch, Phys. Rev. B {\bf 35}, 1851 (1987). https://doi.org/10.1103/PhysRevB.35.1851
\bibitem{Scalettar}R.T. Scalettar, D.J. Scalapino, R.L. Sugar, D. Toussaint, Phys. Rev. B {\bf 39}, 4711 (1989). https://doi.org/10.1103/PhysRevB.39.4711
\bibitem{Staudt}R. Staudt, M. Dzierzawa, A. Muramatsu, Eur. Phys. J. B {\bf 17}, 411 (2000). https://doi.org/10.1007/s100510070120
\bibitem{Jarrell}M. Jarrell, Phys. Rev. Lett. {\bf 69}, 168 (1992). https://doi.org/10.1103/PhysRevLett.69.168
\bibitem{Ulmke}M. Ulmke, V. Jani\v{s}, D. Vollhardt, Phys. Rev. B {\bf 51}, 10411 (1995). https://doi.org/10.1103/PhysRevB.51.10411
\bibitem{Kent}P.R.C. Kent, M. Jarrell, T.A. Maier, T. Pruschke, Phys. Rev. B {\bf 72}, 060411 (2005). https://doi.org/10.1103/PhysRevB.72.060411
\bibitem{Rohringer}G. Rohringer, A. Toschi, A. Katanin, K. Held, Phys. Rev. Lett. {\bf 107}, 256402 (2011). https://doi.org/10.1103/PhysRevLett.107.256402
\bibitem{Hirschmeier}D. Hirschmeier, H. Hafermann, E. Gull, A.I. Lichtenstein, A.E. Antipov, Phys. Rev. B {\bf 92}, 144409 (2015). https://doi.org/10.1103/PhysRevB.92.144409
\bibitem{Kozik}E. Kozik, E. Burovski, V.W. Scarola, M. Troyer, Phys. Rev. B {\bf 87}, 205102 (2013). https://doi.org/10.1103/PhysRevB.87.205102
\bibitem{Vladimir}M.I. Vladimir, V.A. Moskalenko, Theor.\ Math.\ Phys. {\bf 82}, 301 (1990). https://doi.org/10.1007/BF01029224
\bibitem{Metzner}W. Metzner, Phys.\ Rev. B {\bf 43}, 8549 (1991). https://doi.org/10.1103/PhysRevB.43.8549
\bibitem{Pairault}S. Pairault, D. S\'en\'echal, A.-M.S. Tremblay, Eur.\ Phys.\ J. B {\bf 16}, 85 (2000). https://doi.org/10.1007/s100510070253
\bibitem{Sherman18}A. Sherman, J. Phys.: Condens. Matter {\bf 30}, 195601 (2018). https://doi.org/10.1088/1361-648X/aaba0e
\bibitem{Moukouri}S. Moukouri, S. Allen, F. Lemay, B. Kyung, D. Poulin, Y.M. Vilk, A.-M.S. Tremblay, Phys. Rev. B {\bf 61}, 7887 (2000). https://doi.org/10.1103/PhysRevB.61.7887
\bibitem{Rushbrooke}G.S. Rushbrooke, G.A. Baker, P.J Wood, in {\it Phase Transitions and Critical Phenomena}, edited by C. Domb, M.S. Green (Academic Press, New York, 1974).
\bibitem{Sandvik}A.W. Sandvik, Phys. Rev. Lett. {\bf 80}, 5196 (1998). https://doi.org/10.1103/PhysRevLett.80.5196
\bibitem{Holm}C. Holm, W. Janke, Phys. Rev. B {\bf 48}, 936 (1993). https://doi.org/10.1103/PhysRevB.48.936
\bibitem{Kubo}R. Kubo, J. Phys. Soc. Jpn. {\bf 17}, 1100 (1962). https://doi.org/10.1143/JPSJ.17.1100
\bibitem{Abrikosov}A.A. Abrikosov, L.P. Gor’kov, I.E. Dzyaloshinskii, {\it Methods of Quantum Field Theory in Statistical Physics} (Pergamon Press, New York, 1965).
\bibitem{Sherman21}A. Sherman, J. Phys. Soc. Jpn. {\bf 90}, 104707 (2021). https://doi.org/10.7566/JPSJ.90.104707
\bibitem{Sherman24}A. Sherman, J. Low Temp. Phys. {\bf 216}, 800 (2024).
https://doi.org/10.1007/s10909-024-03194-y
\bibitem{Sherman20a}A. Sherman, Eur. Phys. J. B {\bf 93}, 168 (2020). https://doi.org/10.1140/epjb/e2020-10221-4
\bibitem{Sherman23a}A. Sherman, Phys. Scr. {\bf 98}, 065802 (2023). https://doi.org/10.1088/1402-4896/acd281
\bibitem{Sherman23b}A. Sherman, Phys. Scr. {\bf 98}, 115947 (2023). https://doi.org/10.1088/1402-4896/ad000b
\bibitem{Sherman20b}A. Sherman, Phys. Scr. {\bf 95}, 095804 (2020). https://doi.org/10.1088/1402-4896/aba923
\bibitem{Hasenfratz}P. Hasenfratz, Eur. Phys. J. B {\bf 13}, 11 (2000). https://doi.org/10.1007/s100510050003
\bibitem{Press}W.H. Press, S.A. Teukolsky, W.T. Vetterling, B.P. Flannery, {\it Numerical Recipes in Fortran} (Cambridge University Press, Cambridge, 1995), chapter 18.
\bibitem{Jarrell96}M. Jarrell, J.E. Gubernatis, Phys.\ Rept. {\bf 269}, 133 (1996). https://doi.org/10.1016/0370-1573(95)00074-7
\bibitem{Habershon}S. Habershon, B.J. Braams, D.E. Manolopoulos, J.\ Chem.\ Phys. {\bf 127}, 174108 (2007). https://doi.org/10.1063/1.2786451
\bibitem{Slater}J.C. Slater, Phys. Rev. {\bf 82}, 538 (1951). https://doi.org/10.1103/PhysRev.82.538
\end{thebibliography}
\end{document}